\documentclass[aps,pra,twocolumn,showpacs,amsmath,amssymb,floatfix,superscriptaddress,notitlepage]{revtex4-1}

\usepackage{color}
\usepackage{bbm}
\usepackage{graphicx}% Include figure files
\usepackage{dcolumn}% Align table columns on decimal point
\usepackage[utf8]{inputenc}

\usepackage{graphicx}
\usepackage{epstopdf}
\usepackage{amsthm}
\usepackage{amsmath}
\usepackage{empheq}
\usepackage{bbm}
\usepackage{braket}
\usepackage{amssymb}
\usepackage[usenames,dvipsnames]{xcolor}
\usepackage{cases}
\usepackage{latexsym}
\usepackage[colorlinks=true,citecolor=Cerulean,linkcolor=RubineRed,urlcolor=Cerulean]{hyperref}
\usepackage[title]{appendix}

\newtheorem{property}{Property}
\theoremstyle{plain}

\newcommand{\tr}[1]{\operatorname{\textnormal{Tr}}\left[ {#1} \right]}

\newcommand{\be}{\begin{equation}}
\newcommand{\ee}{\end{equation}}

\date{\today}

\begin{document}
\title{Rationally independent free fermions with local hopping}
\author{Jonathon Riddell}
\affiliation{School of Physics and Astronomy, University of Birmingham, Edgbaston, Birmingham, B15 2TT, UK}
\affiliation{School of Physics and Astronomy, University of Nottingham, Nottingham, NG7 2RD, UK}
\affiliation{Centre for the Mathematics and Theoretical Physics of Quantum Non-Equilibrium Systems, University of Nottingham, Nottingham, NG7 2RD, UK}
\author{Bruno Bertini}
\affiliation{School of Physics and Astronomy, University of Birmingham, Edgbaston, Birmingham, B15 2TT, UK}
\affiliation{School of Physics and Astronomy, University of Nottingham, Nottingham, NG7 2RD, UK}
\affiliation{Centre for the Mathematics and Theoretical Physics of Quantum Non-Equilibrium Systems, University of Nottingham, Nottingham, NG7 2RD, UK}

\begin{abstract}
 Rationally independent free fermions are those where sums of single-particle energies multiplied by arbitrary rational coefficients vanish only if the coefficients are all zero. This property guaranties that they have no degeneracies in the many-body spectrum and gives them relaxation properties more similar to those of generic systems. Using classic results from number theory we provide minimal examples of rationally independent free fermion models for every system size in one dimension. This is accomplished by considering a free fermion model with a chemical potential, and hopping terms corresponding to all the divisors of the number of sites, each one with an incommensurate complex amplitude. We further discuss the many-body spectral statistics for these models and show that local probes --- like the ratio of consecutive level spacings --- look very similar to what is expected for the Poisson statistics. We however demonstrate that free fermion models can never have Poisson statistics with an analysis of the moments of the spectral form factor. 
\end{abstract}

\maketitle

\section{Introduction}

In the last two decades there has been a large amount of progress classifying and understanding the dynamics generated by isolated quantum many body Hamiltonians. A rich literature is forming discussing properties of quantum systems in the context of equilibration, thermalization, and information spreading~\cite{PolkovnikovReview, Gogolin_2016, calabrese2016introduction, VidmarRigol, essler2016quench, doyon2020lecture, bastianello2022introduction, alba2021generalized, Ueda2020, fisher2022random}. This also lead us to revisit concepts like quantum chaos, integrability, and localization in the quantum many-body setting~\cite{hayden2007black, sekino2008fast, kos2018many, chan2018solution, Alessio2016, abanin2019many, sierant2024manybody}. 

The field has made several advances in terms of understanding late time dynamics and the corresponding equilibrium values of dynamical quantities such as expectation values, correlation functions, entanglement measures, out of time ordered correlators and more. Traditionally there have been two main approaches for characterising these quantities. The first consists in studying the problem directly in the thermodynamic limit and then explicitly take the limit of large times. This approach is particularly convenient for analytic studies and relies on some special simplifications occurring in the thermodynamic limit~\cite{essler2016quench, fisher2022random}. The second approach instead consists of studying infinite time averages in finite systems. This second approach is more suited to numerical investigations and does not require any special simplification. Instead, late time properties are studied through dephasing arguments based on some typicality assumptions on the spectrum of the system and, sometimes, the functional form of the observables in the energy eigenbasis. Generally one assumes the $q$ no-resonance condition~\cite{Mark2023,Riddell2023,Kaneko2020,riddell2023noresonance,Srednicki_1999}
\begin{equation} 
\label{eq:genericnoresonance}
    \sum_{j\in M} E_j = \sum_{k\in N} E_k\,\, \text{ s.t }\,\, |M|=|N|  \implies M = N,
\end{equation}
where $M,N$ are sets of indices with $q$ entries, identifying the energy values of the model. This allows one to make progress when taking the infinite time average. Such a simple assumption on the spectrum allows one to make strong statements on equilibration and quantum recurrence times \cite{Riddell2023} with fixed $q$ assumptions being the most common in the literature \cite{huang2020instability,Riddell2020,Goldstein2013,Pintos2017,Wilming2018,Herrera2014,chen2023problemtailored,Santos2023,Reimann_2012,Short_2011,Alhambra2020}. Late time properties can typically be deduced with few additional assumptions like eigenstate typicality/thermalization, chaos, large inverse participation ratios etc \cite{Short_2011,Alhambra2020,Huang2017,Yoshida2019,Srednicki_1999}. Eq.~\eqref{eq:genericnoresonance} is a strong assumption on the spectrum and is generally assumed to hold for quantum chaotic models. Recently the spectral statistics of \eqref{eq:genericnoresonance} was studied, and indeed the local spectral statistics for many body chaotic (or random) Hamiltonians for the condition when $q>1$ are Poissonian, indicating that violations of the conditions should be rare or non-existent \cite{riddell2023noresonance}. Probing whether or not a given model violates the condition in Eq.~\eqref{eq:genericnoresonance} is however generically unfeasible, both analytically and numerically.

In this document we focus our attention on a more analytically tractable set of models: free fermionic systems, whose spectral properties have lately been at a centre of a resurgence of interest~\cite{tasaki2010, winer2020exponential,liao2020many, Vidmar2021,lydzba2021single,suntajs2023localization, tasaki2024macroscopic,shiraishi2024nature, camposvenuti2010unitary, camposvenuti2011exact, camposvenuti2013equilibration}. Specifically, we consider free fermion models defined by
\begin{equation}
    \hat{H} = \sum_{n,m=1}^L M_{n,m} \hat{f}_n^\dagger \hat{f}^{\phantom{\dag}}_{m},
\end{equation}
with $M$ Hermitian. One can always solve this model by diagonalising the matrix $M = O \epsilon O^\dagger$, where $O$ is unitary and $\epsilon$ is a diagonal matrix with real entries. Therefore we can write
\begin{equation}
\label{eq:hamgendiag}
    \hat H = \sum_{k=1}^L \epsilon_k d_k^\dagger d^{\phantom{\dag}}_k,
\end{equation}
where $d_k^\dagger = \sum_{n=1}^L O_{n,k} \hat{f}_n^\dagger$.  We make two demands of this model. First we ask it to be extensive, i.e., 
\begin{equation} \label{eq:prop1}
    O_{n,k} = \frac{c_{n,k}}{\sqrt{L}},
\end{equation}
with $c_{n,k} = \mathcal{O}(1)$. The second property we require from our free model is to have a \emph{rationally independent spectrum}, i.e., 
\begin{equation}
\label{eq:prop2}
    \sum_{k=1}^L a_k \epsilon_k = 0,\quad a_k \in \mathbb{Q} \quad \Leftrightarrow \quad  a_k =0, \quad \forall k\,.  
\end{equation}
This property in particular implies that the system cannot have degeneracies in the many particle spectrum. Note that Eq.~\eqref{eq:prop2} also implies the so called single particle  $q$ no resonance condition. Namely 
\begin{equation} 
\label{eq:prop2special}
    \sum_{j \in R_q } \epsilon_{j} =    \sum_{j \in R_q' } \epsilon_{j} \quad \implies \quad  R_q=R_q'\,,
\end{equation}
where $R_q$ and $R_q'$ are two sets of exactly $q$ eigenmodes. Note that the condition in Eq.~\eqref{eq:prop2special} is weaker than the one in Eq.~\eqref{eq:prop2} and does not prevent degeneracies between sectors of the many-particle spectrum characterised by different numbers of particles.

The two properties \eqref{eq:prop1} and \eqref{eq:prop2} imply the free model has exponentially long quantum recurrence times, and the concentration of equilibration is exponentially tight in system size~\cite{Riddell2023}. This is to be compared to the typical tight binding model which has a quantum recurrence time that is linear in size system and a concentration around equilibrium that is a power law \cite{Santos2023,Venuti2013,Carrillo2015,Riddell2020,Kaminishi2015,Kaminishi2019,Gluza2016,Malabarba2015}. Therefore, a model that satisfies Eqs.~\eqref{eq:prop1} and \eqref{eq:prop2} has late time properties more similar to those of a generic system than typically studied free ones. An interesting question is then to find minimal examples of such models and analyse their many-body spectral properties. An important step in this direction has been achieved in Refs.~\cite{tasaki2010, shiraishi2024nature}, where the authors have shown that a simple tight-binding model with complex hopping, and periodic boundary conditions, fulfils \eqref{eq:prop1} and \eqref{eq:prop2special} for prime $L$. More precisely, they show that this is true with probability one when drawing at random the phase of the hopping~\cite{tasaki2010} or for small enough (non-zero) hopping phases~\cite{shiraishi2024nature}. 

In this work we extend this idea to show that if one adds an incommensurate chemical potential then also Eq.~\eqref{eq:prop2} is satisfied. Moreover, we show that if $L$ is not a prime, then both Eqs.~\eqref{eq:prop2} and \eqref{eq:prop2special} are violated. We then prove that Eqs.~\eqref{eq:prop2} and \eqref{eq:prop2special} can be achieved for any $L$ by adding hopping terms corresponding to all the divisors of $L$. We next investigate the spectral properties of the many body spectrum of such models. This is carried out with the ratio test, which showcases results compatible with an underlying Poissonian distribution for the energy levels.Therefore the system seems to follow the Berry Tabor conjecture~\cite{berry1977level} like interacting integrable systems. However, analysing its spectral form factor, and more precisely its higher moments, we show that this is not the case. The spectral statistics is Poissonian only at the level of two point correlations: higher point correlations can never be Poissonian in free systems.

The rest of this manuscript is laid out as follows. In Sec.~\ref{sec:minimalmodel} we construct a class of minimal rationally independent free fermion models and show that they satisfy \eqref{eq:prop1} and \eqref{eq:prop2}. In Sec.~\ref{sec:levelspacing} we show that, contrary to the tight binding model, its level spacing distribution appears Poissonian. In Sec.~\ref{sec:moments} we characterise its spectral statistics using the spectral form factor and its higher moments. Finally, Sec.~\ref{sec:conclusion} contains our conclusions.

\section{Minimal models}
\label{sec:minimalmodel}

In this section we show that to achieve a rationally independent  free fermion model it is sufficient to add a parity-breaking term to the tight binding model. Namely we consider 
\be
\label{eq:hamrealspace}
 \hat{H} = \frac{\alpha + i \beta}{2} \sum_{n=1}^L \hat{f}_n^\dagger \hat{f}^{\phantom{\dag}}_{n+1} +\frac{\gamma}{2} \sum_{n=1}^L \hat{f}_n^\dagger \hat{f}^{\phantom{\dag}}_{n} + {\rm h.c.}.
\ee
In the spin language (i.e.\ via a standard Jordan Wigner transformation) this system is mapped into an XX chain with a magnetic field and a Dzyaloshinskii--Moriya interaction of the form
\be
\propto \sum_j \sigma^{x}_j \sigma^{y}_{j+1} -\sigma^{x}_j \sigma^{y}_{j+1}. 
\ee
 Note that, to avoid complications coming from the boundary terms, here we focus on the spectral properties fermionic Hamiltonian~\eqref{eq:hamrealspace}, rather than the spin chain. Because of the translational invariance of \eqref{eq:hamrealspace}, the diagonal form \eqref{eq:hamgendiag} is simply attained by Fourier transform, i.e.\ by choosing 
\be
\label{eq:Fourier}
 O_{n,k} = \frac{e^{i\frac{2\pi}{L}kj}}{\sqrt{L}}, 
\ee
and gives a dispersion relation of the form  
\begin{equation} \label{eq:hammin}
    \epsilon_k = \alpha \cos\left({\frac{2\pi}{L}k}\right) + \beta \sin\left({\frac{2\pi}{L}k}\right) + \gamma. 
\end{equation}
Note that the $O_{n,k}$ in Eq.~\eqref{eq:Fourier} immediately satisfies our first condition \eqref{eq:prop1}. The validity of \eqref{eq:prop2} is assessed by the following 
\begin{property}
When the couplings $\alpha,\beta$ and $\gamma$ are \emph{incommensurate} the dispersion relation \eqref{eq:hammin} fulfils \eqref{eq:prop2} \emph{iff $L$ is prime}. 
\end{property}
We begin by giving a more precise definition of incommensurate. Let 
\be
\begin{aligned}
&\mathcal {C}_L = \left\{ \cos\left(\frac{2\pi}{L}k_1\right) + \ldots + \cos\left(\frac{2\pi}{L}k_n\right),\right.\\
&\left. \phantom{\left(\frac{2\pi}{L}\right)} n \in \mathbb Z_L,\quad 1\leq k_1<\cdots< k_n \leq L \right\},
\end{aligned}
\ee
be the finite set of points achieved by summing any group of cosine terms generated by our finite quantised wave-numbers. Likewise we denote by $\mathcal{S}_L$ set of sums of sines. We demand that $\alpha,\beta, \gamma$ be chosen such that for any choice of $x \in \mathcal{C}_L, y\in \mathcal {S}_L$ we have that ${x\alpha}/{y\beta}$ and $({x\alpha}+{y\beta})/\gamma$ are irrational. 

This gives us two important properties. First, sums of cosine and sine terms will not cancel each other. Second, sums of different numbers of eigenmodes also cannot be equal. To see how this gives Eq.~\eqref{eq:prop2} we note that, due to the  incommensurability of $\alpha,\beta$ and $\gamma$, we can break the equation on the l.h.s.\ of Eq.~\eqref{eq:prop2} into the following three equations
\begin{align}
    &\displaystyle\sum_{k=1}^{L} a_k\cos\left(\frac{2\pi}{L}k\right)  = 0, \\
    &\displaystyle\sum_{k=1}^{L} a_k\sin\left(\frac{2\pi}{L}k\right) = 0,\\
    &\displaystyle\sum_{k=1}^{L} a_k = 0. 
\end{align}
Focussing on the first two equations and combining them into one we arrive at 
\begin{equation} \label{eq:poly2}
   P(\xi)\equiv \sum_{k = 0}^{L-1} a_k \xi^k = 0,
\end{equation}
where we set  $\xi \equiv \exp\left( i {2 \pi}/{L}\right)$. The third equation is then rewritten as
\be
\label{eq:gammacond}
P(1)=0.
\ee
The key observation now is that Eq.~\eqref{eq:poly2} defines a polynomial with integer coefficients that has a zero at the first $L$-th root of unity. This family of polynomials is well studied in mathematics, with a particularly important role played by the so called $L$-th \emph{cyclotomic polynomial}~\cite{weisstein2002polynomial}, which we denote by $\Phi_L(z)$. The latter is the unique polynomial with roots at  \emph{primitive} $L$-th roots of unity~\footnote{Namely, they are not $N$-roots of unity for some $N<L$.} that is irreducible in $\mathbb Q$, i.e.\ it cannot be factored as the product of two non-constant polynomials with rational coefficients. 

Recalling some basic facts about the cyclotomic polynomial we have that it has degree $\phi(L)\leq L-1$, where $\phi(L)$ is Euler's totient function~\cite{weisstein2002polynomial}. In particular, this means that its degree is strictly smaller than $L-1$ whenever $L$ is not prime. Instead, for prime $L$ one has~\cite{weisstein2002polynomial}
\be
\Phi_L(z) = 1+ z + \cdots + z^{L-1}\,. 
\ee
Since by definition $\Phi_L(z)$ must divide $P(z)$ and the degree of $P(z)$ is at most $L-1$, for $L$ prime we have either $P(z)=0$ or $P(z)=\Phi_L(z)$. Namely we either have $a_k = 0$ for all $k$ or $a_k = 1$ for all $k$. Considering now \eqref{eq:gammacond} we see that it is only fulfilled by the former choice. This proves that, for $L$ prime we have rational independence. 

Next, let us show that Eq.~\eqref{eq:prop2} does not hold for non-prime $L$ by providing an explicit counter example. Consider a system size $L = np$, with $p$ prime and $n>1$, and let $w = \xi^n$. Then $w$ is a $p$-th primitive root of unity and, therefore, it is a zero of the cyclotomic polynomial $\Phi_p(z)$. Namely
\begin{equation}
    \Phi_p(w)=w^{p-1} + w^{p-2} + \ldots 1 = 0. 
\end{equation}
Expressing this equation in terms of $\xi$, we can write it as $Q(\xi)=0$, where we introduced the  polynomial
\begin{equation}
\label{eq:polyQ}
    Q(z) \equiv z^{n(p-1)} + z^{n(p-2)} + \ldots +1.
\end{equation}
Since $Q(\xi) =0$ we can multiply any $R(\xi)$ into $Q(\xi)$ giving $R(\xi) Q(\xi) = 0$. We can use this to construct suitable $P(z)$ fulfilling \eqref{eq:poly2} and \eqref{eq:gammacond}. 

Take for example 
\be
R(z) = z-1,
\ee
and set $P(z)=R(z) Q(z)$. This is a legitimate choice because $R(z) Q(z)$ has degree $L - n + 1 \leq L - 1$. Moreover, since $R(1)=0$ we have that this polynomial also fulfils Eq.~\eqref{eq:gammacond}. \qedsymbol{}

 Note that our proofs shows that also Eq.~\eqref{eq:prop2special} is violated for non-prime $L$. 

Next we move our attention to recover rational independence for composite $L$. To this end we show the following
\begin{property}
To recover rational independence for non-prime $L$ one should add higher harmonics to the dispersion relation. In particular, for a generic $L$ a rationally independent dispersion relation takes the form 
\begin{equation} \label{eq:compH}
     \epsilon_k =  \sum_{d \mid L} \left(\alpha_d \cos\left({\frac{2\pi d}{L} k}\right) + \beta_d \sin\left({\frac{2\pi d}{L} k}\right)\right) + \gamma,
\end{equation}
where the sum is over all the divisors of $L$ strictly smaller than $L$ and $\{\alpha_d, \beta_d\}$, and $\gamma$ are again incommensurable. 
\end{property}
Note that the real space Hamiltonian corresponding to Eq.~\eqref{eq:compH} is recovered by again taking the Fourier transform in Eq.~\eqref{eq:Fourier}. The resulting model can be written similarly to Eq.~\eqref{eq:hamrealspace} with the addition of terms of the form
\begin{equation}
     \hat{H}_d = \frac{\alpha_d + i \beta_d}{2} \sum_{n=1}^L \hat{f}_n^\dagger \hat{f}^{\phantom{\dag}}_{n+d} + {\rm h.c.}.
\end{equation}

To see why this property holds let us consider the case where $L = pq$. First we note that our counter example non-prime $L$ no longer applies if one adds higher harmonics. To see this consider again $L=np$ with $p$ prime and add the term  
\be
\delta \cos\left(\frac{2 \pi p}{L} k\right)+ \eta \sin\left(\frac{2 \pi p}{L} k\right)
\ee
to the dispersion relation \eqref{eq:hammin}. Here $\delta$ and $\eta$ are chosen to retain incommensurability of the individual terms in the dispersion. This means that we now have the following three equations 
\begin{align}
    P(1) = 0, & &P(\xi) = 0, & &P(\xi^p) = 0,
\end{align}
where we again set $\xi=\exp(i 2\pi/L)$. Our choice $P(z)=(z-1)Q(z)$ in the main text, with $Q(z)$ given in Eq.~\eqref{eq:polyQ}, does not satisfy the third equation. Indeed, we have $w^{p} = \xi^{np}= 1$ and $Q(\xi^p)=p\neq 0$. 

Second, we show that when $n$ is also prime we have rational independence if we add harmonics of order $p$ and $n$. This can be seen as follows. Assuming again incommensurability, we obtain that the equation on the l.h.s.\ of Eq.~\eqref{eq:prop2} can be decomposed into the following four equations 
\begin{align}
\label{eq:eqtwoprimes}
    P(1) = 0, & &P(\xi) = 0, & &P(\xi^p) = 0, & & P(\xi^n) = 0.
\end{align}
Considering the last three equations we have that $P(z)$ has zeros at the primitive roots of unity $\xi$, $\xi^p$, and $\xi^n$. This means that it must be proportional to the cyclotomic polynomials $\Phi_{np}(z)$, $ \Phi_{n}(z)$, and $\Phi_{p}(z)$. Namely we have 
\be
 P(z) = R_1(z) \Phi_{np}(z) = R_2(z) \Phi_{n}(z)= R_3(z) \Phi_{p}(z),
\ee
for some $R_j(z)$, $j=1,2,3$. Since by definition each cyclotomic polynomial is irreducible the above equation implies 
\begin{equation}
    P(z) = R_4(z)\Phi_{n}(z) \Phi_{p}(z)\Phi_{np}(z).
\end{equation}
for some $R_4(z)$. 

We now recall another important property of products of cyclotomic polynomials~\cite{weisstein2002polynomial} 
\begin{equation} \label{eq:prodcyc}
    \prod_{d|n} \Phi_d(z) = z^n-1.
\end{equation}
This tells us that $\Phi_{n}(z) \Phi_{p}(z)\Phi_{np}(z)$ is by definition
\begin{equation}
    \Phi_{n}(z) \Phi_{p}(z)\Phi_{np}(z) = 1 + z + \dots + z^{pn-1},
\end{equation}
Since $pn-1$ is the maximal degree that $P(z)$ can have (cf.\ Eq.~\eqref{eq:poly2}) we then conclude $R_4(z) = 0$ or $R_4(z) = 1$. The first choice gives rational independence and the second is incompatible with the first of \eqref{eq:eqtwoprimes}. This concludes the proof. 

Upon adding all $d$ order harmonics such that the step in \eqref{eq:prodcyc} is reproduced, this argument can be directly extended to arbitrary products of powers of primes. Namely, to arbitrary $L\in\mathbb N$.

\section{Level spacing statistics of the many body spectrum}
\label{sec:levelspacing}

In this section we investigate the level spacing statistics of the many body spectrum $E_n = \vec{n} \cdot \mathbf{\epsilon}$. We consider two choices for $\mathbf{\epsilon}$. First we take $\epsilon$ from our minimal model in Eq.~\eqref{eq:hammin}, where $L$ must be prime and we only have nearest neighbour hopping. Second we take a composite $L$ and add the corresponding hopping terms required to produce a rationally independent  free fermion model. In particular we perform the ratio test introduced in Ref.~\cite{Oganesyan2007}. This method avoids unfolding the spectrum while still probing the spectrum's similarity to random matrix theory. Suppose we take $E_n$ and re-order it such that $E_k < E_{k+1}$ for all $k$. Then we define gaps in the spectrum as $s_k = E_k - E_{k-1}$. The ratio test is then concerned with the quantity
\begin{equation} \label{eq:ratio}
		    r_k = \frac{\min \{s_k, s_{k+1} \}}{  \max \{s_k, s_{k+1} \}  }.
		\end{equation}

A spectrum obeying Poisson statistics will follow the distribution~\cite{Atas2013}
\begin{equation} \label{eq:intratio}
		    p(r) = \frac{2}{(1+r)^2}.
		\end{equation}
  and have the mean $\langle r \rangle = 2\ln 2 - 1 \approx 0.38629436112$. 

\begin{figure}[t]
\centering
\includegraphics[width=0.95\linewidth]{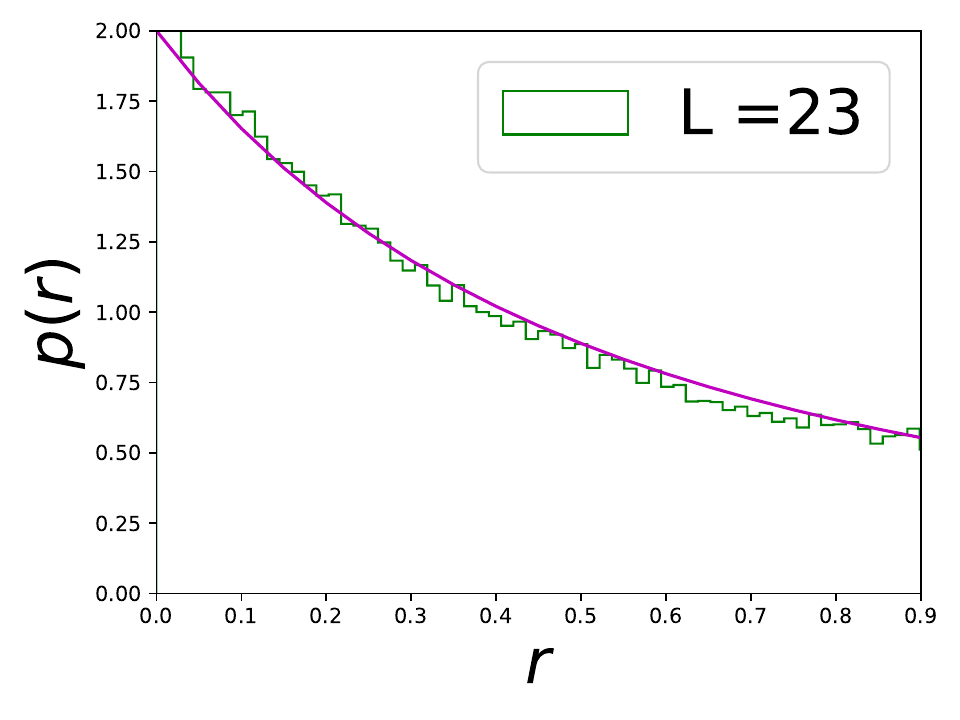}
\includegraphics[width=0.95\linewidth]{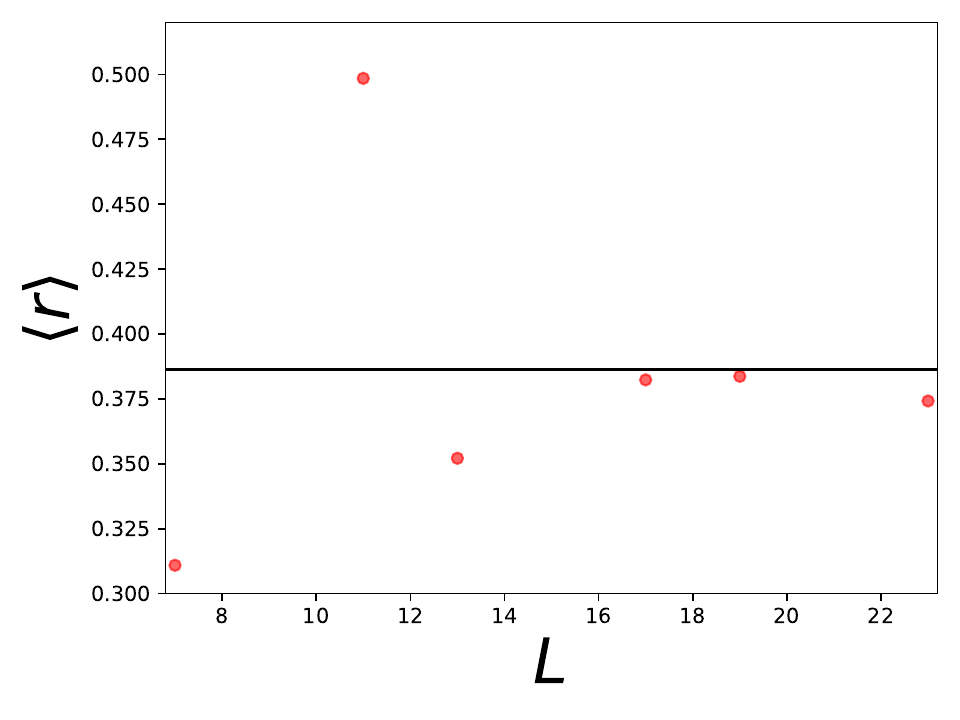}
\caption{In the top figure we have a histogram of the ratio test for the free fermionic system with the dispersion relation $\epsilon_k$ in Eq.~\eqref{eq:hammin} (nearest neighbour hopping) and $L=23$. In purple we plot the Poisson distribution. On the bottom figure we print the average of the ratios for various prime system sizes for the same model. The parameters used are $\alpha = 0.9933$, $\beta = 0.256204$ and $\gamma = 0.452551$.}
\label{fig:ratiotest}
\end{figure}

In Fig.~\ref{fig:ratiotest} we see the surprising result that the nearest neighbour model looks approximately Poisson both when calculating $\langle r \rangle$ and when overlaying its distribution $p(r)$.

\begin{figure}[t]
\centering
\includegraphics[width=0.95\linewidth]{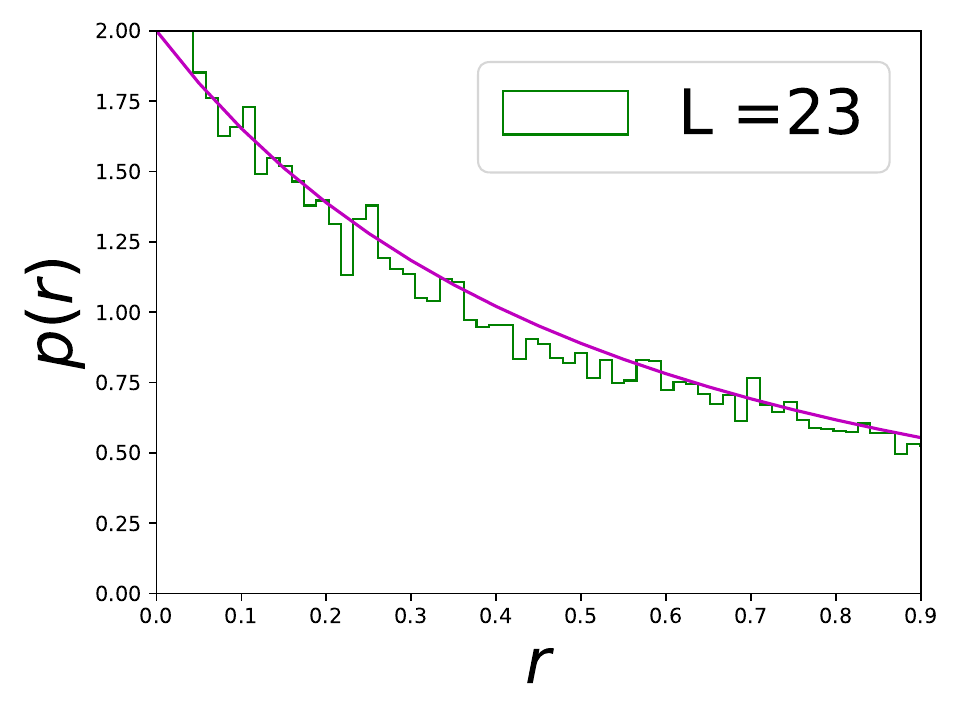}
\caption{In this figure we have a histogram of the ratio test for the free fermionic system with the dispersion relation $\epsilon_k$ in Eq.~\eqref{eq:hammin} (nearest neighbour hopping) and $L=23$ where we have resolved the quasi-momenta and particle number symmetries (see Eq.~\eqref{eq:syms} for details). In purple we plot the Poisson distribution. The parameters used are $\alpha = 0.9933$, $\beta = 0.256204$ and $\gamma = 0.452551$.}
\label{fig:resolvedratiotest2}
\end{figure}

The specific model considered was constructed by randomly generating $\alpha,\beta$ in the interval $[0,1]$ giving us an incommensurate pair. The data presented are obtained for $\alpha = 0.9933$, $\beta = 0.256204$ and $\gamma = 0.452551$, however, we observed that different parameters do not effect the results of Fig.~\ref{fig:ratiotest} for large enough $L$. Namely, one always finds a curve $p(r)$ similar Eq.~\eqref{eq:intratio} along with a mean $\langle r \rangle$ similar to Poisson statistics. The model clearly experiences some form of level attraction but the gaps do not cluster around zero as one expects for a typical tight binding model. At $L = 23$ we observe $\langle r \rangle \approx 0.374197$, a value somewhat similar the expected Poisson result. One may be tempted to conclude that the model has Poisson spectral statistics and is just experiencing relatively slow convergence to that behaviour. We  demonstrate in the next section that this cannot be true by presenting a more extensive probe of the spectral statistics.  Note that to obtain the result in Fig.~\ref{fig:ratiotest} we did not resolve any symmetry of the Hamiltonian. The same statement, however, applies when we resolve a finite number of symmetries, e.g., particle number and translation symmetry. For example, if we only select many body energy eigenvalues that correspond to energy eigenvectors with the properties
\begin{equation} \label{eq:syms}
    \sum_{k=1}^L \langle d_k^\dagger d_k\rangle = \frac{L-1}{2}, \enspace \sum_{k=1}^L k \langle d_k^\dagger d_k\rangle \mod L = 0,
\end{equation}
we find results represented by the one in Fig.~\ref{fig:resolvedratiotest2}. We see that the system continues to look Poissonian according to the ratio test. Specifically, we find $\langle r \rangle \approx 0.37243$, which is also close to the expected Poisson result. This can be expected given the fact that we did not resolve resolve all symmetries as there is an extensive number. Each $d_k^\dagger d_k$ is individually a symmetry of the Hamiltonian.
 
\begin{figure}[t]
\centering
\includegraphics[width=0.95\linewidth]{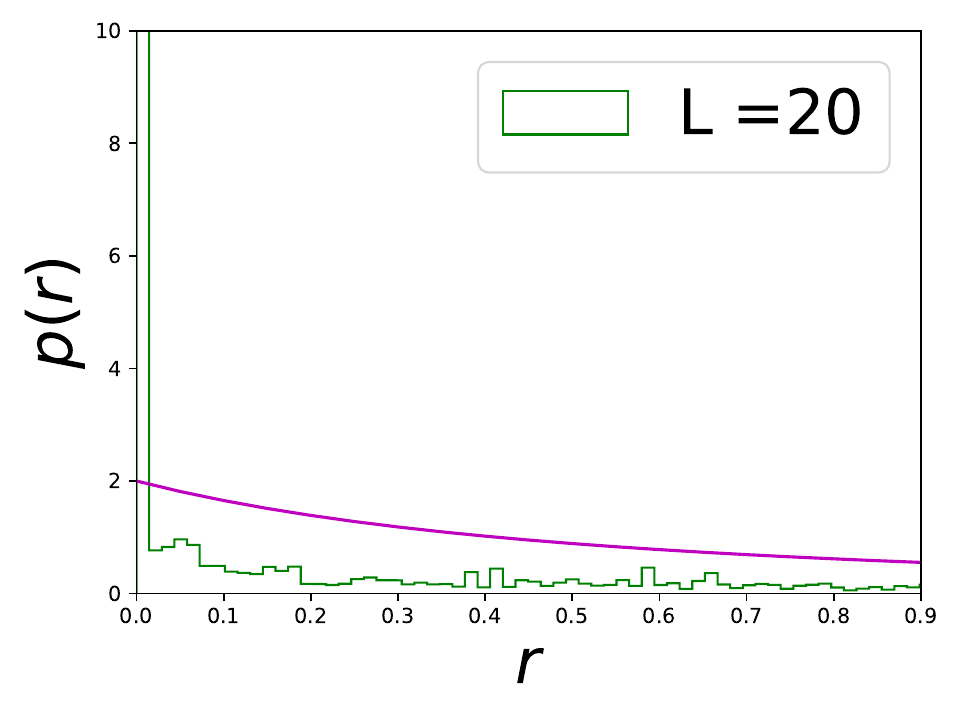}
\caption{In this figure we have a histogram of the ratio test for the free fermionic system with the dispersion relation $\epsilon_k$ in Eq.~\eqref{eq:hammin} (nearest neighbour hopping) and $L=20$. In purple we plot the Poisson distribution. The parameters used are $\alpha = 0.9933$, $\beta = 0.256204$ and $\gamma = 0.452551$.}
\label{fig:ratiotest2}
\end{figure}

For comparison we plot the ratio test for $L = 20$ and ~\eqref{eq:hammin}, a non-prime system size, in Fig.~\ref{fig:ratiotest2}. The ratios cluster around zero with a large amount of exact degeneracies. 
Note, in the case of the tight binding model, the vast majority of gaps are zero and we would have an approximate delta function at $0$. 

\begin{figure}[t]
\centering
\includegraphics[width=0.95\linewidth]{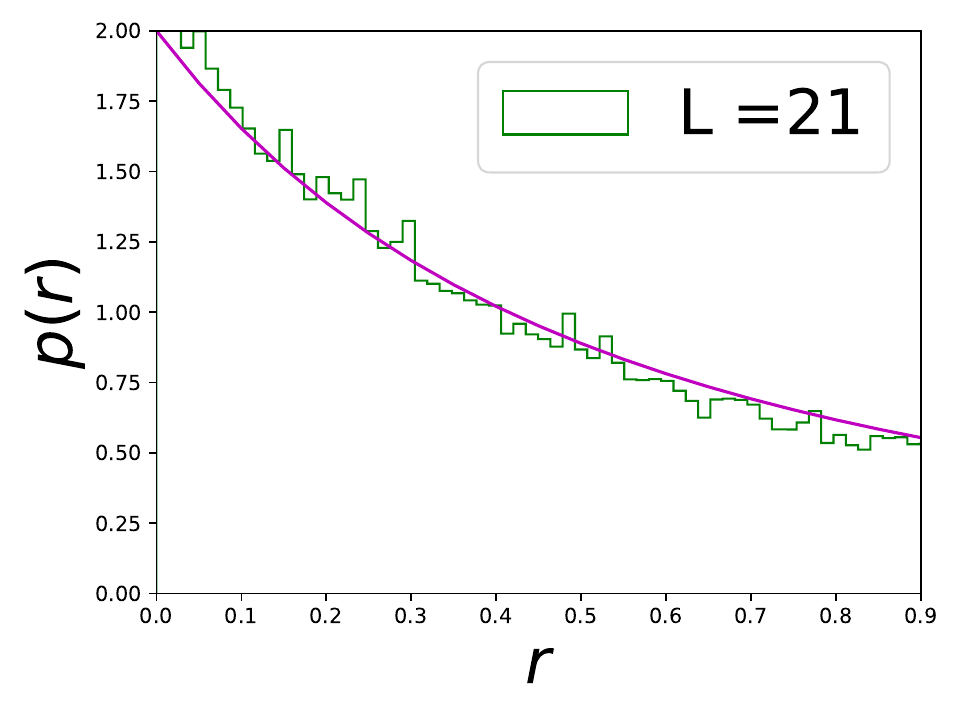}
\caption{In this figure we have a histogram of the ratio test for the free fermionic system with the dispersion relation $\epsilon_k$ in Eq.~\eqref{eq:compH} (sum of different hopping terms) and $L=21$. In purple we plot the Poisson distribution. The parameters used are $(\alpha_1,\beta_1,\alpha_3,\beta_3,\alpha_7,\beta_7,\gamma) \approx (0.9933,0.2562,0.4525,0.0772,0.6086,0.8501,0.5201)$.}
\label{fig:compratiotest2}
\end{figure}

 Finally we demonstrate that adding in the appropriate hopping terms for composite $L=21$ (from Eq.~\eqref{eq:compH} we must have hopping terms for $d=1,3,7$), we once again recover a distribution for the ratio test consistent with Poisson statistics. This is shown in Fig.~\ref{fig:compratiotest2}. Here we find $\langle r \rangle \approx 0.375196$. In this case, the parameters used in the figure are $(\alpha_1,\beta_1,\alpha_3,\beta_3,\alpha_7,\beta_7,\gamma) \approx (0.9933,0.2562,0.4525,0.0772,0.6086,0.8501,0.5201)$.
\section{Moments of the spectral form factor}
\label{sec:moments}

A more robust probe of the spectral statistics is given by the so called spectral form factor (SFF)~~\cite{kos2018many, chan2018spectral, bertini2018exact, friedman2019spectral, suntajs2020quantum, bertini2021random, moldgalya2021spectral, suntajs2021spectral, garratt2021local, garratt2021manybody, bertini2022exactspectral, shivam2023many, chan2022many, suntajs2023localization}, which is the Fourier transform of the spectral two-point function. The SFF is expressed as 
\begin{equation} 
\label{eq:SFF}
    K(t,L) = \mathbb E[\left|\tr{ \left(U_t\right) }\right|^{2}],
\end{equation}
where $U_t = e^{i \hat{H} t}$ and $\mathbb E[\cdot]$ is an average (either over an ensemble of similar systems or in time) that removes non-universal features~\cite{prange1997the}. Contrary to the level spacing statistics this quantity tests spectral correlations over all energy scales, not only over short ones. 

To probe also higher point correlations among energy levels, here we consider arbitrary higher moments of the SFF. Namely  
\begin{equation} 
\label{eq:SFFq}
    K_q(t,L) = \mathbb E[\left|\tr{ \left(U_t\right) }\right|^{2q}],
\end{equation}
and take $\mathbb E[\cdot]$ to be the moving time-average over a time window of size $\tau$ 
\begin{equation}
    \mathbb E[f(t)] = \frac{1}{\tau} \int_{t}^{t+\tau} f(s) {\rm d}s.
\end{equation}
In fact, to make the expressions more readable we will always consider the limit of large $\tau$~\footnote{More precisely we take $t\gg \tau \gg 1$.}.

Let us now compute $K_q(t,L)$ in our minimal rationally independent  free fermionic model and compare our results with the predictions of the Poissonian statistics, which gives the following SFF higher moments for large $L$~\cite{haake2001quantum} 
\begin{equation} 
\label{eq:poiss}
    {K_q(t,L)} =   2^{qL} q!.
\end{equation}
We begin by considering $q=1$ and noting that the trace can be expressed as
\begin{equation}
\tr{ \left(U_t\right) } = \sum_{m=1}^{2^L} e^{it \vec{m}\cdot \vec \epsilon},
\end{equation}
where $\vec{m}$ is the $L$-bit binary representation of $m$ and ${\vec \epsilon}= ( \epsilon_1, \ldots, \epsilon_L)$. This means that we can write  
\begin{equation}
\left| \tr{ \left(U_t\right) }\right|^2 = \sum_{m,n} e^{it(\vec{m}-\vec{n})\cdot \vec \epsilon}.
\end{equation}

A term survives the time average and the $\tau\to\infty$ limit iff
\begin{equation}
    (\vec{m}-\vec{n})\cdot \vec \epsilon = 0.
\end{equation}
This is clearly fulfilled when $\vec m = \vec n$. To see that this is the only case we expand it as 
\begin{equation}
    (\vec{m}-\vec{n})\cdot \vec \epsilon=\sum_k (m_k-n_k)\epsilon_k = 0.
\end{equation}
$m_k-n_k \in \mathbb{Z}$. It then follows that our only choice is for $m_k - n_k = 0$ because of the rational independence of $\epsilon_k$.  Putting all together we then have 
\begin{equation}
     {K_1(t,L)} = \prod_{k=1}^L \sum_{m_k} 1 = 2^L.
\end{equation}
Therefore our spectrum gives the first moment expected for the Poisson ensemble.

Considering now arbitrary higher moments we have 
\begin{equation}
\label{eq:qpower}
    \left|\tr{ \left(U_t\right) }\right|^{2q} = \sum_{m_{i},n_{i}=1}^{2^L} e^{i t \mathbf{\epsilon} \cdot\sum_{i=1}^q \left(\vec{m}_{i} - \vec n_{i} \right) }.
\end{equation}

We now wish to understand which terms survive time average and infinite $\tau$ limit. This is again when the argument of the exponential is zero
\begin{equation}
    \mathbf{\epsilon} \cdot\sum_{i=1}^q \left(\vec{m}_{i} - \vec{n}_{i} \right) = 0, 
\end{equation}
writing this component-wise gives
\begin{equation}
    \sum_{k=1}^L \epsilon_k \sum_{i=1}^q \left( m_{ik} - n_{ik} \right) = 0. 
\end{equation}
The number $\sum_{i=1}^q \left( m_{ik} - n_{ik} \right)$ takes integer values in $[-q,q]$. Therefore, the rational independence of $\{\epsilon_k\}$ tells us that the only solution to this equation is 
\begin{equation}
    \sum_{i=1}^q \left( m_{ik} - n_{ik} \right) = 0,\qquad \forall k\,.
\end{equation}

Fixing a set of $m_{ik}$ tells us that there are precisely 
\be
\binom{q}{\sum_i m_{ik}}, 
\ee
choices for $n_{ik}$ to satisfy this equation. This is because we can take $\{n_{ik}\}_{i=1}^q$ to be any permutation of $\{m_{ik}\}_{i=1}^q$ but the permutations only differing by a reshuffling of the ones $(\sum_i m_{ik}!)$ or the zeroes $((q-\sum_i m_{ik})!)$ have to be counted only once. Likewise fixing $ \sum_i m_{ik}$ gives us 
\be
\binom{q}{\sum_i m_{ik}}
\ee 
total configurations of $m_{ik}$. Plugging this back in the average of Eq.~\eqref{eq:qpower}, summing over possible values of $ \sum_i m_{ik}$, and using the identity 
\be
\sum_{k=0}^q \binom{q}{k}^2 = \binom{2q}{q},
\ee
this gives 
\begin{equation} \label{eq:freemom}
    {K_q(L)} = \prod_{k=1}^L \binom{2q}{q} =  \binom{2q}{q}^L.
\end{equation}
This result agrees with Eq.~\eqref{eq:poiss} for $q = 1$ but disagrees with it for $q>1$. Note in particular that the disagreement comes because Eq.~\eqref{eq:freemom} is larger than Eq.~\eqref{eq:poiss}. This happens because, even though the Hamiltonian in Eq.~\eqref{eq:hamrealspace} has a non-degenerate spectrum for $\alpha, \beta, \gamma$ incommensurate and $L$ prime, the spectrum of 
\be
\hat H_q=\sum_{j=0}^{q-1} \overbrace{I \otimes \cdots  \otimes I}^j \otimes \hat H\otimes I\otimes \cdots \otimes  I,
\ee
has far more degeneracies than 
\be
\hat H_{{\rm int},q}=\sum_{j=0}^{q-1} \overbrace{I \otimes \cdots  \otimes I}^j \otimes  \hat H_{\rm int}\otimes I\otimes \cdots \otimes  I\,,
\ee
where $\hat H_{\rm int}$ is an Hamiltonian with a generic Poisson distributed spectrum, e.g.\ that of a Bethe Ansatz integrable model. Indeed, whereas $\hat H_{{\rm int},q}$ has an obvious permutation symmetry, $\hat H_q$ is invariant under permutations of each separate mode. Therefore, $e^{i H}$ behaves as the tensor product of $L$ Poissonian matrices (and indeed has the same ${K_q(L)}$, see, e.g., Ref~\cite{bertini2022exactspectral}). This means that free fermionic models cannot display Poisson statistics in the many body spectrum. However, thanks to the rational independence of $\{\epsilon_k\}$, one can claim that our model produces the smallest possible SFF moments for a free fermionic system.

\section{Conclusion}
\label{sec:conclusion}

In this article we have identified a simple family of free fermionic systems for which we could prove rational independence of the single-particle spectrum, i.e., that for fixed system size its energy eigenvalues are linearly independent over rational numbers. Free fermionic systems with this property are known to have special late time behaviour including ``interacting-like" concentration of equilibration and quantum recurrence times. 

We have also shown that this family of models has a non-degenerate many body spectrum that under local probes of the spectrum (the ratio test~\cite{Oganesyan2007}) looks strikingly Poissonian --- this is not the case when rational independence is lifted as seen in Fig.~\ref{fig:ratiotest}. Studying the higher moments of the spectral form factor, however, we have shown that the spectral statistics of our systems agrees with the Poissonian prediction only at the level of two-point spectral correlations. In fact, this is the closest any free system can ever be to have Poissonian statistics. This shows that the predictions of local probes should be taken with care. 

An immediate direction for future research is to identify other dynamical signatures of rationally independent single-particle spectrum in the context of local interactions. In particular, a fascinating question concerns the interplay between rational independence and interacting perturbations. Since rational independence reduces the number of thermalising channels --- for a fixed size there are no non-trivial energy conserving scattering processes --- systems with rational independence should anomalously long prethermalization plateaus in finite volume. 

\begin{acknowledgments}
We thank Maurizio Fagotti and Hosho Katsura for useful discussions.  We thank Henrik Wilming and Lorenzo Campos Venuti for bringing Refs.~\cite{shiraishi2024nature} and \cite{tasaki2010} to our attention. We also thank Lorenzo Campos Venuti for pointing out a mistake in Eq.~\eqref{eq:freemom} of the first version of this paper. We acknowledge financial support from the Royal Society through the University Research Fellowship No.\ 201101 and warmly acknowledge the hospitality of the Simons Center for Geometry and Physics during the program ``Fluctuations, Entanglements, and Chaos: Exact Results'' where part of the project has been carried out.
\end{acknowledgments}

\bibliography{bibliography.bib}

\end{document}